\documentclass{article}

\usepackage{arxiv}

\usepackage[utf8]{inputenc} 
\usepackage[T1]{fontenc}    
\usepackage{hyperref}       
\usepackage{url}            
\usepackage{booktabs}       
\usepackage{amsfonts}       
\usepackage{nicefrac}       
\usepackage{microtype}      
\usepackage{amsmath,amsthm,amssymb} 
\usepackage{cleveref}       
\usepackage{lipsum}         
\usepackage{graphicx}
\usepackage[Export]{adjustbox}
\usepackage{tabularx}
\usepackage{cellspace}
\usepackage{bbm}
\usepackage{doi}
\usepackage{subcaption}
\usepackage{algpseudocode}
\usepackage{algorithm}
\usepackage{bm}

\usepackage{datetime}
\settimeformat{hhmmsstime}

\usepackage{todonotes}
\usepackage[binary-units=true]{siunitx}

\title{Statistical Reconstruction For Anisotropic X-ray Dark-Field Tomography}

\newif\ifuniqueAffiliation

\ifuniqueAffiliation 
\author{ \href{https://orcid.org/0000-0000-0000-0000}{\includegraphics[scale=0.06]{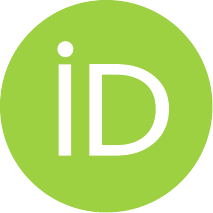}\hspace{1mm}David~Frank}
    Department of Computer Science \\
    School of Computation, Information and Technology\\
    Technical University of Munich \\
	\texttt{frankd@cit.tum.de} \\
	\And
	\href{https://orcid.org/0000-0000-0000-0000}{\includegraphics[scale=0.06]{orcid.pdf}\hspace{1mm}Tobias Lasser}\thanks{What should I do with NVIDIA affiliation?} \\
    School of Computation, Information and Technology\\
    Technical University of Munich \\
	\texttt{lasser@cit.tum.de}
}
\else
\usepackage{authblk}

\newbox{\orcid}\sbox{\orcid}{\includegraphics[scale=0.06]{orcid.pdf}} 
\author[1,2]{%
	\href{https://orcid.org/0009-0006-4989-2144}{\usebox{\orcid}\hspace{1mm}David Frank\thanks{\texttt{frankd@cit.tum.de}}}%
}
\author[1]{
    \href{https://orcid.org/0000-0002-3441-5648}{\usebox{\orcid}\hspace{1mm}Cederik Höfs\thanks{\texttt{cederik.hoefs@tum.de}}}%
}
\author[1,2]{%
	\href{https://orcid.org/0000-0001-5669-920X}{\usebox{\orcid}\hspace{1mm}Tobias Lasser\thanks{\texttt{lasser@cit.tum.de}}}%
}
\affil[1]{School of Computation, Information and Technology, Technical University of Munich, Germany}
\affil[2]{Munich Institute of Biomedical Engineering, Technical University of Munich, Germany}
\fi

\renewcommand{\shorttitle}{Statistical Reconstruction for Anisotropic X-ray Dark-Field Tomography}

\hypersetup{
pdftitle={Statistical Reconstruction for Anisotropic X-ray Dark-Field Tomography},
pdfsubject={q-bio.NC, q-bio.QM},
pdfauthor={David Frank, Cederic Höfs, Tobias Lasser},
pdfkeywords={First keyword, Second keyword, More},
}

\algrenewcommand\algorithmicrequire{\textbf{Input:}}
\algrenewcommand\algorithmicensure{\textbf{Output:}}

\begin{document}
\maketitle

\begin{abstract}
Anisotropic X-ray Dark-Field Tomography (AXDT) is a novel imaging technology that enables the extraction of fiber structures on the micrometer scale, far smaller than standard X-ray Computed Tomography (CT) setups. Directional and structural information is relevant in medical diagnostics and material testing. Compared to existing solutions, AXDT could prove a viable alternative. Reconstruction methods in AXDT have so far been driven by practicality. Improved methods could make AXDT more accessible. We contribute numerically stable implementations and validation of advanced statistical reconstruction methods that incorporate the statistical noise behavior of the imaging system. We further provide a new statistical reconstruction formulation that retains the advanced noise assumptions of the imaging setup while being efficient and easy to optimize. Finally, we provide a detailed analysis of the optimization behavior for all models regarding AXDT. Our experiments show that statistical reconstruction outperforms the previously used model, and particularly the noise performance is superior. While the previously proposed statistical method is effective, it is computationally expensive, and our newly proposed formulation proves highly efficient with identical performance. Our theoretical analysis opens the possibility to new and more advanced reconstruction algorithms, which in turn enable future research in AXDT.
\end{abstract}


\section{Introduction}

X-ray computed tomography (CT) revolutionized medical diagnostics. However, unlike other imaging modalities, such as magnetic resonance imaging, where Diffusion Tensor Imaging (DTI) \cite{basser1995InferringMicrostructuralFeaturesa} can extract directional structures, no such information can be extracted from the sample using attenuation-based X-ray CT. Methods such as DTI have provided invaluable insights into the inner workings of the human brain. A review of applications related to DTI is provided in \cite{assaf2008DiffusionTensorImagingb}.

However, X-ray CT is typically superior to MRI-based techniques in imaging speed, resolution and availability. Classical attenuation-based X-ray CT cannot image directional information and lacks the resolution to obtain fine-grained structures such as white brain matter. Nevertheless, using complementary effects of the X-ray wave such as refraction and (ultra) small-angle scattering, one can extract structures smaller than the resolution of the detector and hence microstructure information of the sample. Studies suggest that the small-angle scattering, commonly referred to as dark-field signal, provides valuable Information in the diagnostic and staging of lung diseases \cite{guo2024GratingbasedXrayDarkfield,frank2022DarkfieldChestXray,scherer2017XrayDarkfieldRadiography, modregger2016SmallAngleXray,yaroshenko2015ImprovedVivoAssessment,velroyen2015GratingbasedXrayDarkfield}.

The dark-field signal can be extracted using grating-based interferometry. Initially developed for high brilliance X-ray sources\cite{weitkamp2005XrayPhaseImaging}, it was extended to standard X-ray sources \cite{pfeiffer2006PhaseRetrievalDifferential,pfeiffer2008HardXrayDarkfieldImaginga}. This makes grating-based tomography an excellent candidate for many applications. For example, \cite{viermetz2022DarkfieldComputedTomography} integrated a grating-interferometry setup into a standard clinical CT gantry, enabling X-ray Dark-field CT on the human scale.

This grating-based X-ray setup only captures scattering perpendicular to the grating orientation. To capture directional information, the sample must be measured from multiple non-standard positions. The directional information can be extracted using the novel imaging modality Anisotropic X-ray Dark-Field Tomography (AXDT) \cite{wieczorek2016AnisotropicXRayDarkFielda,wieczorek2017MicrostructureOrientationExtraction}. It enables the reconstruction of scattering strengths and directions in arbitrary directions for each voxel. Unlike DTI, AXDT promises to provide fast scan times and higher resolution. \cite{wieczorek2018BrainConnectivityExposed} provided evidence that AXDT can successfully expose the orientation of nerve fiber connections in the human brain.

As AXDT is still a novel technology, one seeks reconstruction techniques that allow for enhanced reconstruction quality, which in turn enable the reduction of the induced X-ray dose. Both are required for adoption in material testing and biomedical use cases. So far, all research conducted on AXDT has used a simplified linearized model for reconstruction. One would expect this to systematically overestimate and, as such, introduce a bias \cite{fessler2016statisticalrecosntruction}.
\cite{schilling2017StatisticalModelsAnisotropic} presented reconstruction formulations for AXDT that make correct noise assumptions for the grating-based imaging process. However, no numerically stable implementation has been proposed until now, and as such, no detailed analysis of these models could be conducted.

We contribute such a numerically stable implementation for the statistical reconstruction formulation provided by \cite{schilling2017StatisticalModelsAnisotropic}. Further, we showcase its capabilities on two relevant experimental datasets, and provide a detailed analysis of the model's optimization behavior including important tools for optimization such as bounds on the Lipschitz constants, which is a fundamental tool to ensure convergence of algorithms such as the Fast gradient method \cite{kim2016OptimizedFirstorderMethods}.

Additionally, we provide a new problem formulation that simplifies the statistical reconstruction, while retaining the same noise assumptions. This model is more efficient and easier to optimize than the previous model, which makes statistical reconstruction more approachable. Further, unlike the model provided by \cite{schilling2017StatisticalModelsAnisotropic}, we can provide bounds on the Lipschitz constant as well.

In the next section, we introduce the grating-based X-ray CT setup used for AXDT. AXDT itself is explained in detail in the subsequent section. We follow this up with a detailed discussion of both the linearized problem formulation and the statistical approaches. Next, we discuss the extraction of the fiber orientation from an AXDT reconstruction. We conclude the methods section with a derivation of the computation of the Lipschitz constants for the problem formulations and a short overview of optimization techniques used throughout our experiments. In Chapter \ref{sec:experiments}, we describe our experiments in detail. We follow this up with a discussion of the results in Chapter \ref{sec:discussion} and a conclusion in Chapter \ref{sec:conclusion}.

\section{Methods}
\label{sec:methods}

\subsection{Related Work}
\label{sec:related-work}

\subsection*{Grating-based Tomography}
\label{sec:grating-tomo}

\begin{figure}[h]
	\centering
	\includegraphics[width=\linewidth]{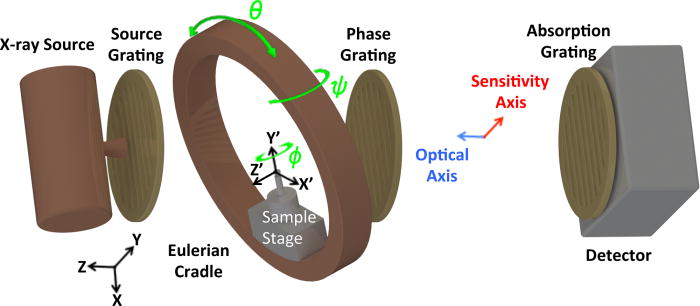}
	\caption{Schematic overview of grating-based interferometry setup used in Anisotropic X-ray Dark-Field Tomography. The setup consists of an X-ray source, the three gratings (source, phase and absorption), and a sample stage mounted on an Euler cradle. This setup enables acquiring non-standard acquisition viewpoints needed in AXDT. Figure by \cite{sharma2016SixDimensionalXray} is licensed under \href{https://creativecommons.org/licenses/by/4.0/}{CC BY 4.0}}
	\label{fig:gbi-setup}
\end{figure}

Grating-based tomography, as introduced by \cite{weitkamp2005XrayPhaseImaging,pfeiffer2006PhaseRetrievalDifferential,pfeiffer2008HardXrayDarkfieldImaginga},  relies on the Talbot-interferometry to generate fringe patterns on the detector. These fringe patterns can be generated with a typical X-ray setup augmented with 3 gratings, as depicted in Figure \ref{fig:gbi-setup}. Changes in the fringe pattern enable the extraction of absorption, dark-field and phase information of the sample.

To extract these signals, the absorption grating is moved in small increments perpendicular to the optical axis and in-plane of the gratings for each viewpoint. This series captures variations in the interference pattern. For each pixel, this series creates a sinusoidal pattern, which enables the extraction of absorption, differential phase-contrast, and dark-field images \cite{pfeiffer2008HardXrayDarkfieldImaginga}. To capture the anisotropic behavior of the dark-field signal and reconstruct directional information of the sample, non-standard trajectories are used \cite{sharma2018advanced}. In the case of grating-based tomography, an Euler cradle, as shown in Figure \ref{fig:gbi-setup} can be used.

The absorption is related to the mean intensity change of the sinusoidal pattern in relation to a reference scan. The dark-field relates to the visibility of the interference pattern, particularly the change in amplitude. As X-rays scatter in the sample, the interference pattern is blurred, reducing its contrast. As such, the visibility is a direct measure of the dark-field signal.

Given the intensity signal $I_j(x_g)$, the $j$-th measurement, and phase stepping position $x_g$, the intensity can be expressed as a truncated Fourier series \cite{pfeiffer2008HardXrayDarkfieldImaginga}:
$$I_j(x_g) \approx a_{j} + b_{j} \cos(kx_g + \phi_{j,1})$$
with $a_{j}$ and $b_j$ the amplitude coefficients, $\phi_{j,i}$ the corresponding phase coefficients, $k = \frac{2\pi}{p_2}$ and $p_2$ the period of G2. The transmission can be retrieved using an additional reference signal as
$$T_j = \frac{a_{j}^s}{a_{j}^r}$$
with the superscripts $s$ and $r$ denote the signal with the sample and the reference signal, respectively.

Scattering information is contained in the amplitude of the second Fourier component $b_{j}$. It decreases when X-rays are scattered or reflected at internal inhomogeneities and interfaces on their passage through the specimen. This results in an image formed exclusively from higher-angle diffraction intensities scattered by the specimen. The relative decrease of the visibility can be quantified by the normalized visibility.
$$d_j = \frac{b_{j}^s a_{j}^r}{a_{j}^s b_{j}^r}$$
We refer to $d_j$ as the dark-field signal.

\subsection*{Anisotropic Dark-Field Tomography}
\label{sec:axdt}

The dark-field signal is inherently anisotropic, which means that it depends on the relative orientation of the sample with the propagation of the X-rays and the grating orientation. In order to reconstruct three-dimensional scattering functions, the measurements must be well distributed over all possible scattering orientations \cite{sharma2018advanced}.

\cite{wieczorek2016AnisotropicXRayDarkFielda} defined a closed-form continuous forward model for AXDT. The model can be discretized using spherical harmonics and enables an elegant and efficient way to perform the reconstruction for AXDT. We will denote the linear operator associated with it as $\mathcal{B}$.

The least squares formulation is a typical formulation to minimize and reconstruct inverse problems. With the forward model in place, we can use this problem formulation to obtain the discretized spherical-harmonic coefficients $\eta_m^l$ and thus enable the extraction of the scattering magnitude $\eta$. This is the first model to reconstruct AXDT with the explicit formulation:
$$f_{m1}(\eta) = \frac{1}{2}||\mathcal{B}\eta + \ln(d)||^2_2$$
We will refer to this model as \textit{m1}. We can reformulate it similar to \cite{schilling2017StatisticalModelsAnisotropic}
$$f_{m1}(\eta) = \mathbbm{1}^T \left(\mathcal{B}\eta + \ln(d)\right)^2$$
where $\mathbbm{1}^T$ is defined as the row vector of ones and gives us a shorthand notation for the sum.
To analyze the statistical properties of this model, we define the random vector $D = (D_j)$, where $D_j$ is a random variable to the $j$-th dark-field coefficient. The model \textit{m0} assumes that $\ln(D)$ is Gaussian, or at least close to Gaussian. As mentioned above, one would expect this to systematically overestimate and, as such, introduce a bias \cite{fessler2016statisticalrecosntruction}. Still to date, all research conducted in the field of AXDT used this model for the reconstruction.

To find a solution, one can then optimize:
$$ \underset{\eta}{\arg\min} \quad f_{m1}(\eta)$$
Many algorithms exist to reconstruct this specific problem. As the problem is differentiable, the class of First-Order methods is a common choice. Accelerated methods based on Nesterov's fast gradient method exist \cite{kim2016OptimizedFirstorderMethods}. This class of algorithms profits from an approximation of the Lipschitz constant to pick a suitable step size to ensure convergence. Newton-Raphson algorithms additionally use the Hessian matrix. However, the exact computation of the Hessian is often expensive, and as such, the Quasi-Newton methods use an approximation of the Hessian to avoid the expensive computation of the Hessian. A well-established Quasi-Newton algorithm is the Broyden–Fletcher–Goldfarb–Shanno (BFGS) algorithm \cite{broyden1970ConvergenceClassDoubleranka,fletcher1970NewApproachVariable,goldfarb1970FamilyVariablemetricMethods,shanno1970ConditioningQuasiNewtonMethods}. An important low-memory variant of the algorithm was developed \cite{liu1989limited}. This variant stores only the previous $m$ iterations of values of $y$ and $s$ to approximate the Hessian of $f$.

Another algorithm that is popular in the X-ray CT community, is the method of conjugate gradients (CG) \cite{Hestenes1952,shewchuk1994introduction}. The classical method solves a linear system of equations with a positive definite matrix. The Popular least squares formulations can be solved with the algorithm via the normal equation. The algorithm chooses the search directions in such a way that the descent directions are conjugate to each other. In practice, this results in an algorithm that has excellent convergence. Extension to non-linear functions exists. Unlike the linear CG, the step size for each step cannot be chosen automatically, as a line search algorithm is necessary for the non-linear extension of CG.

Given the reconstructed spherical harmonics coefficients, we use the algorithm described in \cite{wieczorek2017MicrostructureOrientationExtraction} to extract the fiber orientation and magnitude, which enables us to visualize the microstructures of the given samples. The algorithm is based on the Funk-Radon Transform. The extracted local scattering function is transformed into an orientation density function using the Funk-Radon Transform. The local maxima of the orientation density functions contain information about the strongest fiber orientation for each voxel. Recently, \cite{huyge2023FiberOrientationEstimation} proposed a different algorithm for fiber extraction, which is based on constrained spherical deconvolution. The authors show that for reinforced polymers, their algorithm improves the fiber extraction compared to the Funk-Radon approach. Here, we rely on the previous algorithm to draw comparison to previously published research based on the Funk-Radon Transform. Further, no comparisons for biomedical samples is given for the approached based on constrained spherical deconvolution.

\subsection{Our Contribution}
\label{sec:contribution}

\subsection*{Numerical Stable Statistical Reconstruction for AXDT}
\label{sec:stats-reco}

In the linear model, the negative logarithm is applied to the dark-field signal $d$ and the reconstruction is performed on $-ln(d)$. When applying statistical reconstruction, the original measurements are used as a concrete realization of random variables. With this, one can incorporate nonlinear physical effects. In contrast to the linear model, one aims to maximize the log-likelihood. Here, we formulate the optimization problems as a minimization of the negative log-likelihood to stay consistent with the linear model.

In a grating-based system, \cite{chabior2011SignaltonoiseRatioRay} showed  that the extracted amplitudes of the Fourier approximation have to follow the distributions:
$$ A_j \sim \mathcal{N}\left(a_j, \frac{a_j}{N}\right), \quad \frac{1}{2}B_j \sim \mathcal{R}\left(\frac{b_j}{2}, \sqrt{\frac{a_j}{2N}}\right)$$
Capital letters $A_j$ and $B_j$ are random variables corresponding to the extracted $a_j$ and $b_j$; $\mathcal{N}$ denotes the Gaussian distribution and $\mathcal{R}$ the Rician distribution. For a statistical reconstruction, \cite{schilling2017StatisticalModelsAnisotropic} derived a log-likelihood function. Formulating it as a minimization problem, it is defined as:
$$f_{m2}(\mu, \eta) = \mathbbm{1}^T \left( \frac{3}{2}\ln(a_\mu) + \frac{N\left(2a + 2a_\mu^2 + b^2 + (\alpha a_\mu d_\eta)^2\right)}{4a_\mu} - \ln I_0\left(\frac{N}{2}b\alpha d_\eta\right) \right)$$
where $I_0$ denotes the modified Bessel Function of the first kind, $\alpha = \frac{b^r}{a^r}$ the visibility of the reference scan, $N$ the number of phase-steps performed, and $a_\mu = \exp(-\mathcal{A}\mu)$ and $d_\eta = \exp(\mathcal{B}\eta)$ the forward projected values of $\mu$ and $\eta$ respectively. \cite{schilling2017StatisticalModelsAnisotropic} referred to this model as \textit{m2b}, we refer to it as \textit{m2}. This model accurately describes the statistical behavior of the AXDT reconstruction. They further derived the first and second derivatives of the loss function. However, naive implementations are numerically unstable due to the modified Bessel functions. In the following, we provide insights into the necessary steps to implement the formulation in a numerically stable way.

The $k$-th modified Bessel function of the first kind is defined as
$$I_k(t) = \frac{1}{\pi} \int_0^\pi \exp(t \cos x) \cos(kx)dx$$
Notice that it contains the exponential function, which leads to numerical issues in the reconstruction process of AXDT. However, in the evaluation of the loss function, it is only ever used inside the logarithm. As such, we can use an implementation of $\ln I_0$, which eliminates the exponential and thus makes the evaluation numerically stable.

Further, the gradient of $f_{m2}$ with respect to $\eta$ and the Hessian contain a division of the modified Bessel Function of first order by the zeroth order. This is again numerically challenging, if the implementation evaluates $I_1(x)$ and $I_0(x)$ individually. \cite{schilling2017StatisticalModelsAnisotropic} already notes that $\frac{I_1(x)}{I_0(x)}$ quickly goes to \num{1} for $x \to \infty$. Using this, we directly compute $\frac{I_1(x)}{I_0(x)}$, instead of each individually, for the gradient and Hessian of the model. This resolves the numerical instabilities and enables reconstruction of AXDT with statistically accurate noise models.

\subsection*{Simplified Statistical Reconstruction}
\label{sec:new-model}

Model \textit{m2} reconstructs values for scattering and attenuation at the same time. For AXDT, we care most about the reconstruction of $\eta$, making the output $\mu$ of the model \textit{m2} superfluous for our use cases. The model, as shown in our experiments below, is computationally expensive and difficult to optimize. As such, we propose a new model, derived from \textit{m2}, which we refer to as \textit{m3}. The new model is based on the Rician noise assumption of the measurements; however, it assumes all parts associated with the attenuation to be known and constant. As such, the noise assumption simplifies to
$$\frac{1}{2}B_j \sim \mathcal{R}\left(\frac{1}{2}b_j, \sqrt{\frac{\tilde{a}}{2N}}\right)$$
Similarly to model \textit{m2}, we can derive a loss function as a negative log-likelihood based on this assumption:
$$f_{m3}(\eta) = \mathbbm{1}^T \left( \frac{N}{4} a\alpha^2d_\eta^2 - \ln I_0\left(\frac{N}{2} b\alpha d_\eta \right) \right)$$
The gradient with respect to $\eta$ of \textit{m3} is defined as:
$$ \nabla f_{m3}(\eta) = \mathcal{B}^*\left(z \frac{I_1(z)}{I_0(z)} - \frac{N}{2}a\alpha^2d_\eta^2\right)$$
with $z = \frac{N}{2}b\alpha d_\eta^2$ and $\mathcal{B}^*$ denoting the adjoint operation of the AXDT forward model. Finally, the Hessian is defined as
$$H_{m3}(\eta) = \mathcal{B}^* \operatorname{diag}\left(N a\alpha^2d_\eta^2 + z^2 \left(\left(\frac{I_1(z)}{I_0(z)}\right)^2 - 1\right)\right)\mathcal{B}$$
\cite{schilling2017StatisticalModelsAnisotropic} showed that \textit{m2} is locally convex. Using their results, it's straight forward to show that model \textit{m3} is similarly locally convex. Model \textit{m3} retains the noise assumptions of the most general model, but is a simpler function. In particular, it is computationally less expensive and easier to optimize, which enables faster reconstruction with similar quality.

\subsection*{Lipschitz Condition for the Loss Functions}
\label{sec:lipschitz-cond}

Many interesting reconstruction algorithms, such as FGM require $L$-smooth functions to provide convergence guarantees. In this section, we will derive bounds on Lipschitz constants for the models \textit{m1} and \textit{m3}.

A function $f$ is $L$-smooth if and only if the spectral norm of the Hessian is bounded by $L$ \cite[Chapter~5]{Beck2017}:
$$ f \quad L\text{-smooth} \Leftrightarrow ||H_f(x)|| \le L \quad \forall x$$
Further, physics dictate that $a < a_\mu \Leftrightarrow \mathcal{A}\mu \ge 0$ and $d \le 1 \Leftrightarrow \mathcal{B}\eta \ge 0$. As such we restrict our domain to the respective sets.

The spectral norm of the AXDT operator $\mathcal{B}$ is bounded by:
$$
	|| \mathcal{B} || \le \frac{1}{4\pi} \sum_{k=0}^K\sum_{m=-k}^k || W_k^m \mathcal{A} ||
	\le \frac{1}{4\pi} \sum_{k=0}^K ||\mathcal{A}||
	\le \frac{K}{4\pi} ||\mathcal{A}|| \\
$$
with the fact $W_k^m < 1$, which is given from the discretized forward model for AXDT as described by \cite{wieczorek2016AnisotropicXRayDarkFielda}.

The spectral norm is the square root of the largest Eigenvalue of an operator. As such, we can use the power iterations to approximate the spectral norm of $\mathcal{A}$. As model \textit{m1} is based on the least squares formulation, the Lipschitz constant can be bound by:
$$L_{m1} = ||\mathcal{B}^*\mathcal{B}|| \le \left(\frac{K}{4\pi} ||\mathcal{A}||\right)^2$$

For model \textit{m3}, we use the \cite[p.~794]{Eriksson2004} states that for a diagonal matrix $A = \operatorname{diag}(a_i)$, the spectral norm is
$$||A|| = \max(|a_i|)$$ and apply this to the diagonal part of the Hessian. This leads to:
\begin{align*}
	||H_{m3}|| & \le ||\mathcal{B}||^2 \max \left(\left|N a \alpha^2 d_\eta^2 - z^2\left(1 - \frac{I_1(z)}{I_0(z)}\right)\right|\right)^2 \\
	           & \le ||\mathcal{B}||^2N \max (a \alpha^2 + b \alpha)
\end{align*}

This leads us to the bounds on the Lipschitz constant for \textit{m3}:
$$ L_{m3} \le  N \max (a \alpha^2 + b \alpha) \left(\frac{K}{4\pi} ||\mathcal{A}||\right)^2$$
We have yet to find a suitable bound for the formulation of model \textit{m2}.

\section{Experiments}
\label{sec:experiments}

\paragraph{Samples} To evaluate the reconstruction quality and convergence behavior of the noise models, we use two different datasets. The first sample consists of two crossed wooden sticks. The sample was also used in \cite{wieczorek2016AnisotropicXRayDarkFielda}. The sample was measured with a symmetrical x-ray grating interferometer with an integrated distance of \SI{91}{\cm}. Two absorption gratings with \SI{10}{\micro\m} periods were used, and a \SI{5}{\micro\m} period phase grating. X-rays were generated with a tungsten-target X-ray tube run at \SI{60}{kVp} and \SI{13.3}{\milli\ampere} current. The spectrum was filtered by \SI{2}{\mm} aluminum. A Varian PaxScan 2520D flat-panel detector with a CsI scintillator and 800x800 pixels of size \SI{127}{\mm^2} was used to record images. For each position, a series of eight phase-stepping images were acquired with 1s exposure time each, from which amplitudes of the Fourier approximation are extracted. In total, 1200 different viewpoints sampling the unit sphere were acquired.

The second sample is a cerebellum of a human brain. The same setup was used as in the crossed sticks. The X-ray tube operated at \SI{1.66}{\milli\ampere}. The acquisition had 1404 viewpoints, again with 8 phase-stepping images acquired at each position with a 2 s exposure time. The sample was previously published in \cite{wieczorek2018BrainConnectivityExposed}

\paragraph{Hardware} The machine used for reconstruction has an AMD EPYC 7452 \num{32}-Core (\num{64} Threads) CPU running at \num{2.35} GHz with \qty{515820}{\mebi\byte} RAM and \num{6} NVIDIA Quadro RTX 6000 (\qty{24576}{\mebi\byte} VRAM). The machine is running Ubuntu 22.04 with the Linux kernel 5.15.0-91-generic and using the NVIDIA driver version 545.23.08 with CUDA version 12.3.

All reconstructions are performed using elsa \cite{frank2024ElsaElegantFramework} version 0.9.0. The samples are binned by a factor of \num{4} resulting in an image size of $160^2$ pixels per measurement. The volume is set to size $160^3$ with a spacing of $0.127$ in each direction. The weighting matrices $W_k^m$ forming a part of the forward model are precomputed using the weighting function as \cite{wieczorek2017MicrostructureOrientationExtraction} $h(u, t, l) \mapsto (|l\times u| |u, t|)^2$. The truncation degree of $K = 4$ is used similar to previous work \cite{wieczorek2016AnisotropicXRayDarkFielda}, as well as the restriction to even degrees.

\begin{figure}[!htbp]
	\centering
	\begin{subfigure}{\linewidth}
		\includegraphics[width=\linewidth]{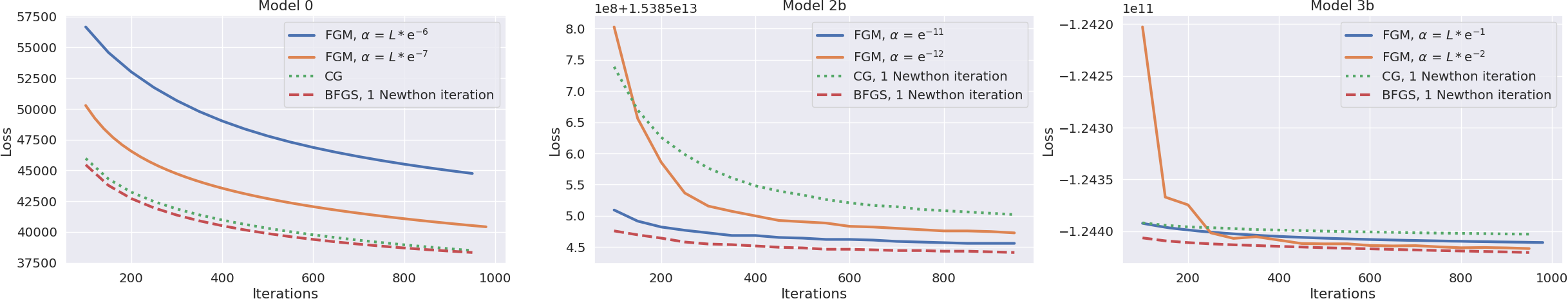}
		\label{fig:convergence_loss}
	\end{subfigure}
	\begin{subfigure}{\linewidth}
		\includegraphics[width=\linewidth]{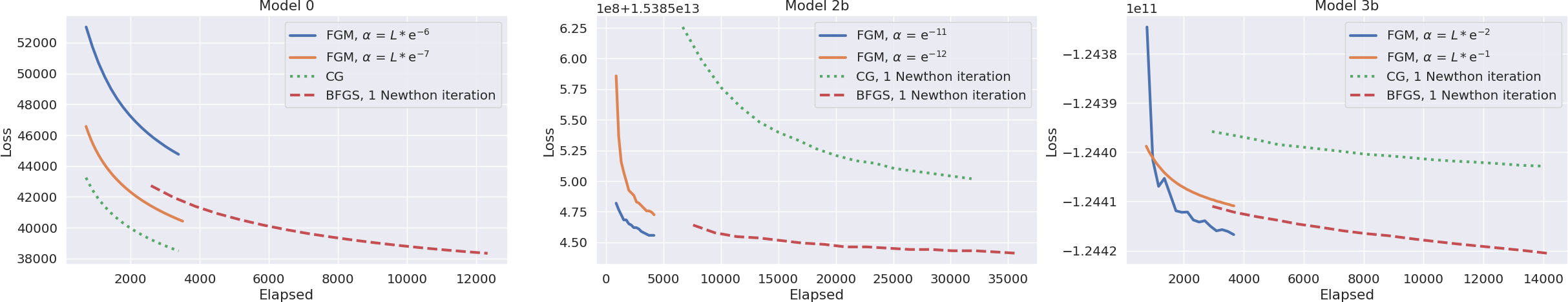}
		\label{fig:convergence_time}
	\end{subfigure}
	\caption{Convergence plot for reconstruction of the crossed sticks sample. Top: plot of loss over number of iterations. Bottom: Loss over time plots.}

	\label{fig:convergence}
\end{figure}

\paragraph{Reconstruction} We performed the reconstruction for each noise model with CG, FGM and L-BFGS. For the linear model \textit{m1} we used Conjugate gradients for solving the Least Squares (CGLS) \cite{Hestenes1952} as a numerically stable and efficient version of CG. For the non-linear models, we used the non-linear extension of CG with Newton-Raphson and Barzilai Borwein as line search algorithms. The reconstructions using L-BFGS use Newton-Raphson as a line search. We investigated a couple of different numbers of iterations for Newton-Raphson for the line search. In our experiments, more than a single iteration of the did either not result in better reconstruction or often resulted in non-converging experiments. As such, the results shown in Figure \ref{fig:convergence} are restricted to a single iteration of Newton-Raphson. As Barzilai Borwein has no guarantees on decrease of the function, it is not suited for L-BFGS and in our experiments it did not lead to any successful reconstructions. L-BFGS has a parameter for the amount of memory used; we've also investigated the influence of this parameter. However, the influence on convergence is minimal in our testing; as such, we stick to 10 saved results to approximate the Hessian. Finally, FGM requires a fixed step length in the interval from $(0, \frac{2}{L})$; for all but model \textit{m2} we have a bound on $L$, as such we choose multiple values in the interval and evaluate their convergence. For model \textit{m2}, we choose fixed step lengths in a uniform interval. The convergence results for the crossed sticks sample can be seen in Figure \ref{fig:convergence}. We show the spherical harmonic coefficients for ta slice of the crossed wooden stick sample in Figure \ref{fig:visu_sticks_harmonics}.

\paragraph{Fiber extraction} For fiber extraction, we chose the reconstruction from L-BFGS for the nonlinear models. For model \textit{m1}, we chose CG, even though L-BFGS converged slightly faster; previous work used CG, and  as such, we want to compare our research to the previously used algorithm. We verified that the reconstructions with similar losses have a small mean square error. We extracted the fibers at different iterations and chose the visually best results for the final visualization. The output is visualized using ParaView. For the crossed sticks sample, see Figure \ref{fig:visu_sticks}. And see Figure \ref{fig:visu_brain} for the visualization of the brain sample. For the summary of the number of iterations used in the figures and runtime for each model, see Table \ref{table:reco_time}.

Table \ref{table:sticks_scatter} shows the mean, variance and \SI{95}{\percent} quantile for the scattering strength of the reconstruction of the crossed wooden stick sample. We compute the values once for the compelte reconstruction volume, and then again for a segmentation containing only the sticks.

\begin{table}[!h]
	\centering
	\begin{tabular}{ccccccc}
		\toprule
		{}          & \multicolumn{3}{c}{Crossed Sticks sample} & \multicolumn{3}{c}{Brain sample}                                                                        \\ \cmidrule(lr){2-4} \cmidrule(lr){5-7}
		{}          & Iterations                                & Time                             & Time per      & Iterations & Total Time              & Time per      \\
		Model       & {}                                        & (hh:mm:ss)                       & iteration (s) & {}         & (hh:mm:ss)              & iteration (s) \\
		\midrule
		\textit{m1} & $240$                                     & \formattime{00}{13}{13}          & $3.31$        & $20$       & \formattime{00}{01}{13} & $3.69$        \\
		\textit{m2} & $1800$                                    & \formattime{18}{14}{23}          & $36.48$       & $280$      & \formattime{03}{32}{56} & $45.63$       \\
		\textit{m3} & $240$                                     & \formattime{01}{00}{50}          & $15.21$       & $60$       & \formattime{00}{16}{49} & $16.83$       \\
	\end{tabular}
	\caption{Summary of number of iterations, time and time per iteration for the reconstructions visualized in Figure \ref{fig:visu_sticks} and Figure \ref{fig:visu_brain}.}
	\label{table:reco_time}
\end{table}

\begin{table}[!h]
	\centering
	\begin{tabular}{*7c}
		\toprule
		{}          & \multicolumn{3}{c}{Complete Volume} & \multicolumn{3}{c}{Segmentation of Sticks}                                                                                                          \\ \cmidrule(lr){2-4} \cmidrule(lr){5-7}
		Model       & Mean                                & Variance                                   & q95                    & Mean                   & Variance                    & q95                    \\
		\midrule
		\textit{m1} & $1.975\mathrm{e}^{-4}$              & $3.020\mathrm{e}^{-8}$                     & $5.435\mathrm{e}^{-4}$ & $7.142\mathrm{e}^{-4}$ & $1.050\mathrm{e}^{-7}$      & $1.294\mathrm{e}^{-3}$ \\
		\textit{m2} & $1.170\mathrm{e}^{-4}$              & $\bm{1.221\mathrm{e}^{-8}}$                & $2.763\mathrm{e}^{-4}$ & $6.595\mathrm{e}^{-4}$ & $\bm{7.925\mathrm{e}^{-8}}$ & $1.141\mathrm{e}^{-3}$ \\
		\textit{m3} & $1.021\mathrm{e}^{-4}$              & $\bm{8.683\mathrm{e}^{-9}}$                & $2.183\mathrm{e}^{-4}$ & $6.603\mathrm{e}^{-4}$ & $\bm{8.715\mathrm{e}^{-8}}$ & $1.156\mathrm{e}^{-3}$ \\
	\end{tabular}
	\caption{Mean, variance and \SI{95}{\percent} quantile of the reconstruction scattering strength for the reconstruction of the crossed sticks sample depicted in Figures \ref{fig:visu_sticks_harmonics} and \ref{fig:visu_sticks}. Values computed for both the complete reconstructed volume and a segmentation to only the sticks.}
	\label{table:sticks_scatter}
\end{table}

\section{Discussion}
\label{sec:discussion}

\begin{figure}[!tbhp]
	\setlength\tabcolsep{6pt}
	\adjustboxset{width=\linewidth,valign=c}
	\centering
	\begin{tabularx}{1.0\linewidth}{@{}
			l
			X @{\hspace{2pt}}
			X @{\hspace{2pt}}
			X
			@{}}
		\rotatebox[origin=c]{90}{Order 0, Rank 0}
		 &
		\begin{subfigure}[t]{\linewidth}
			\includegraphics{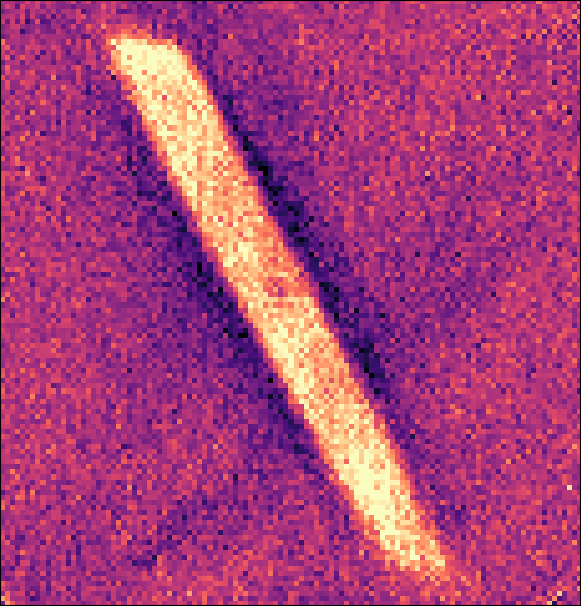}
		\end{subfigure}
		 &
		\begin{subfigure}[t]{\linewidth}
			\includegraphics{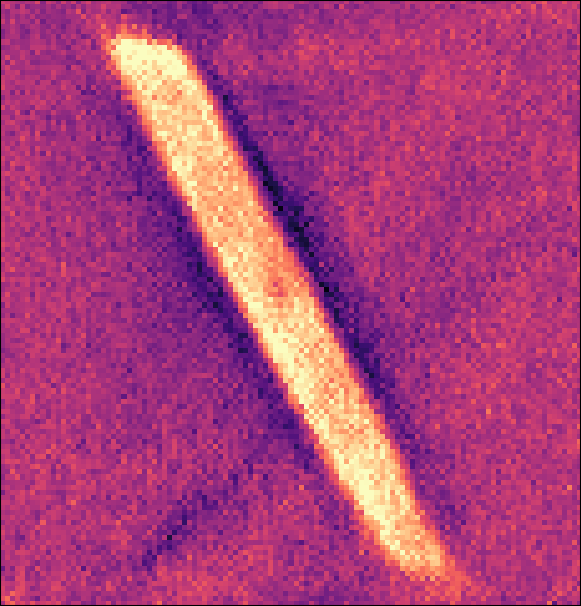}
		\end{subfigure}
		 &
		\begin{subfigure}[t]{\linewidth}
			\includegraphics{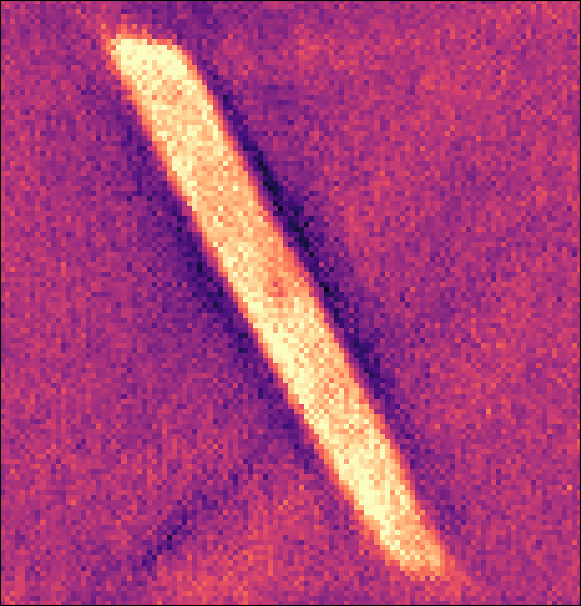}
		\end{subfigure} \vspace{0.1pt} \\
		\rotatebox[origin=c]{90}{Order 2, Rank 1}
		 &
		\begin{subfigure}[t]{\linewidth}
			\captionsetup{justification=centering}
			\includegraphics{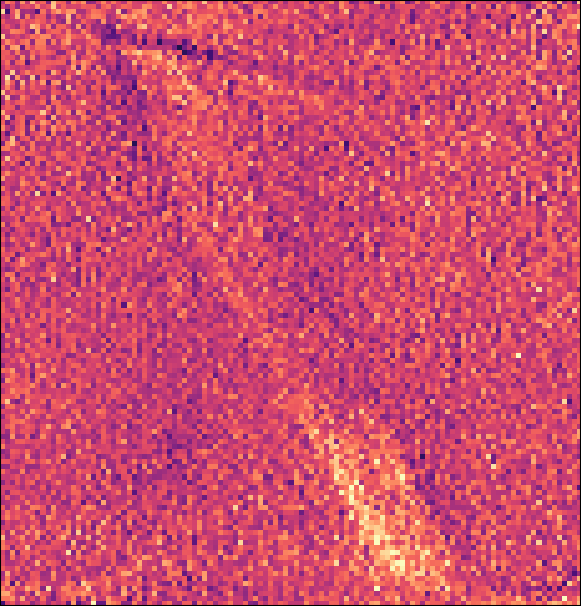}
			\caption{Spherical Harmonics coefficients of model \textit{m1}}\label{fig:visu_sticks_harmonics_m1}
		\end{subfigure}
		 &
		\begin{subfigure}[t]{\linewidth}
			\captionsetup{justification=centering}
			\includegraphics{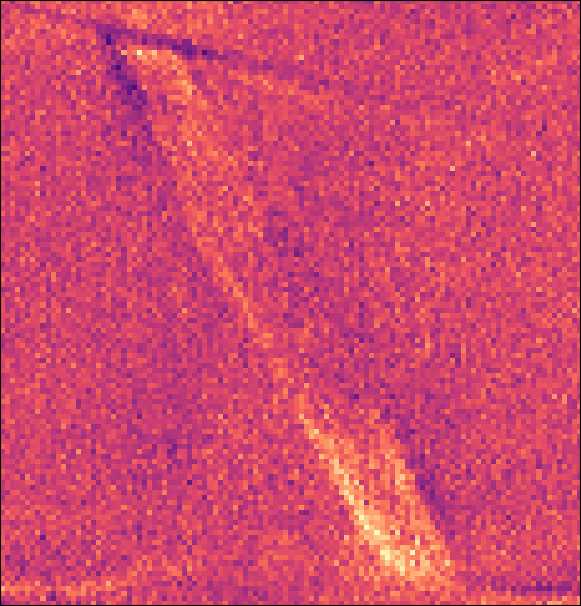}
			\caption{Spherical Harmonics coefficients of model \textit{m2}}\label{fig:visu_sticks_harmonics_m2}
		\end{subfigure}
		 &
		\begin{subfigure}[t]{\linewidth}
			\captionsetup{justification=centering}
			\includegraphics{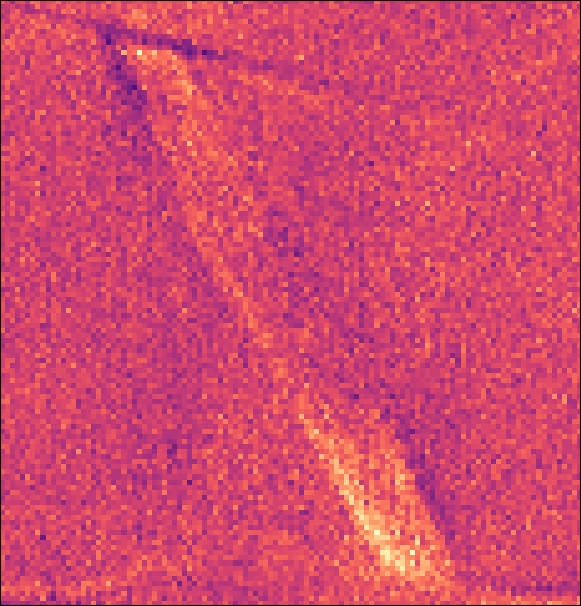}
			\caption{Spherical Harmonics coefficients of model \textit{m3}}\label{fig:visu_sticks_harmonics_m3}
		\end{subfigure}
	\end{tabularx}
	\caption{Visualization of slice 68 of the spherical harmonic coefficients $\eta_l^m$ of the crossed wooden sticks, emphasizing the improved noise behavior of the statistical reconstruction methods. The top row visualizes $\eta_0^0$ and the bottom one $\eta_2^1$. From left to right, coefficients of the models \textit{m0}, \textit{m2}, \textit{m3} respectively. For each row, the images are windowed to the same interval.}
	\label{fig:visu_sticks_harmonics}
\end{figure}

\begin{figure}[!htbp]
	\centering
	\begin{subfigure}[t]{0.33\linewidth}
		\includegraphics[width=\linewidth]{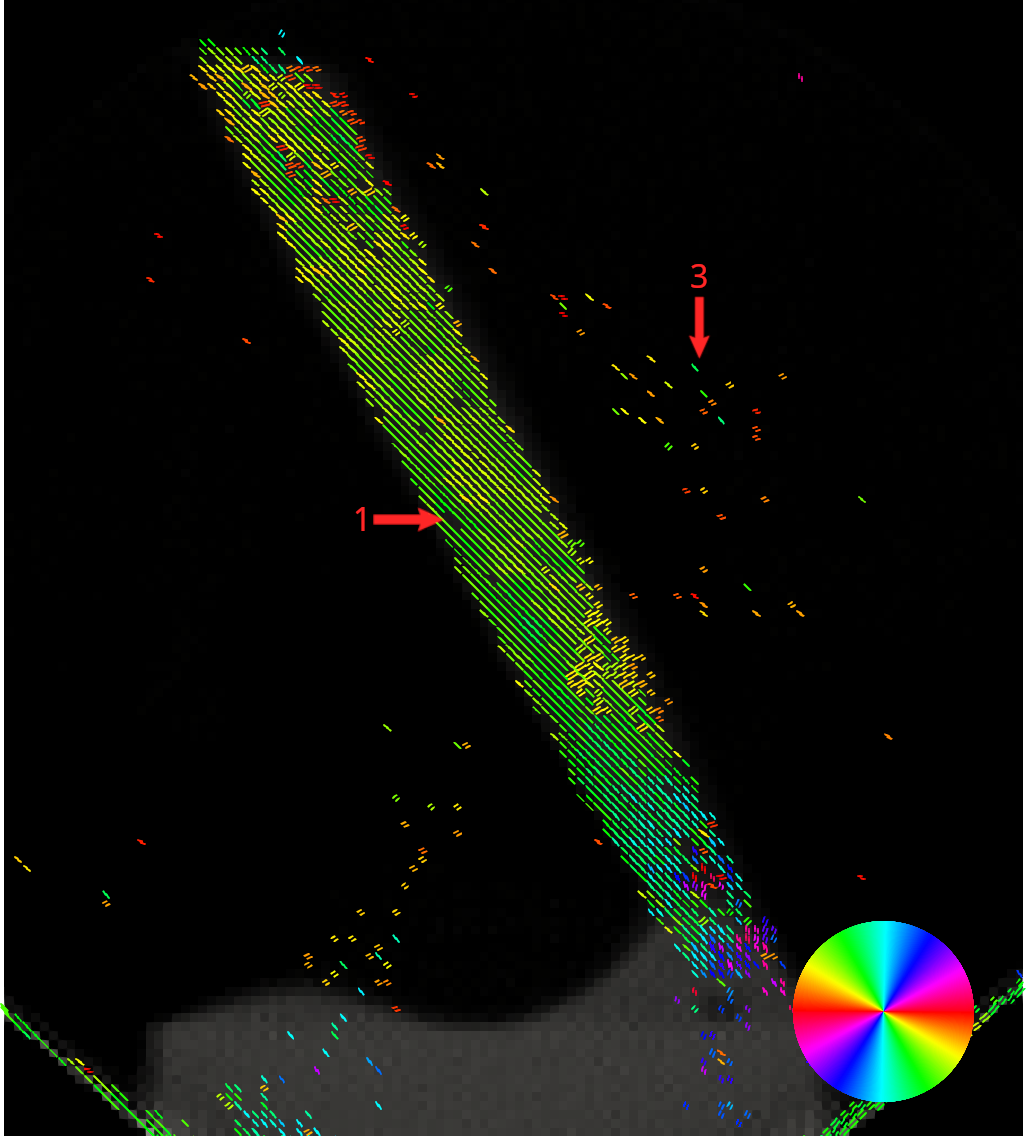}\hfil%
		\includegraphics[width=\linewidth]{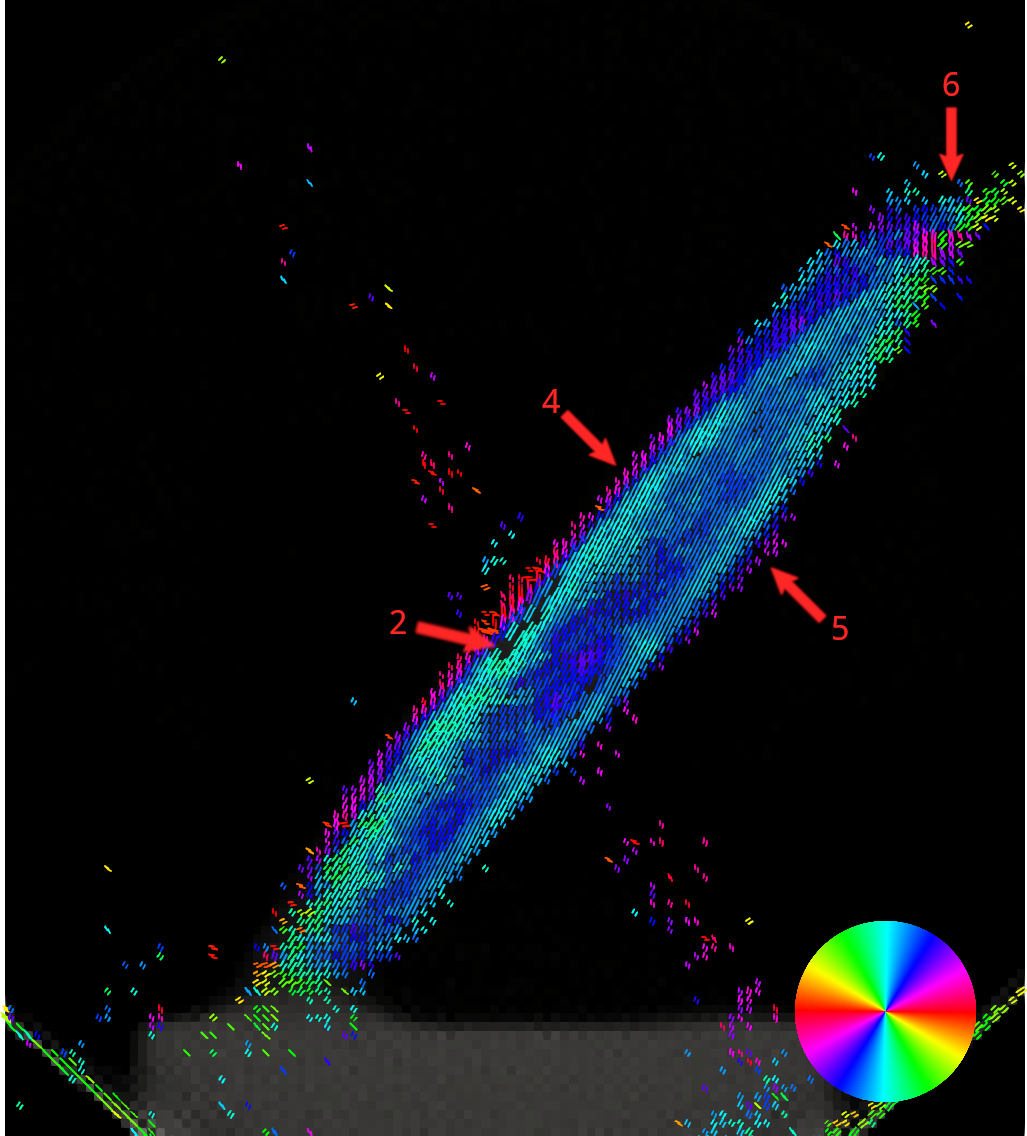}
		\caption{Visualization of model \textit{m1}}\label{fig:visu_sticks_m0}
	\end{subfigure}\hfil%
	\begin{subfigure}[t]{0.33\linewidth}
		\includegraphics[width=\linewidth]{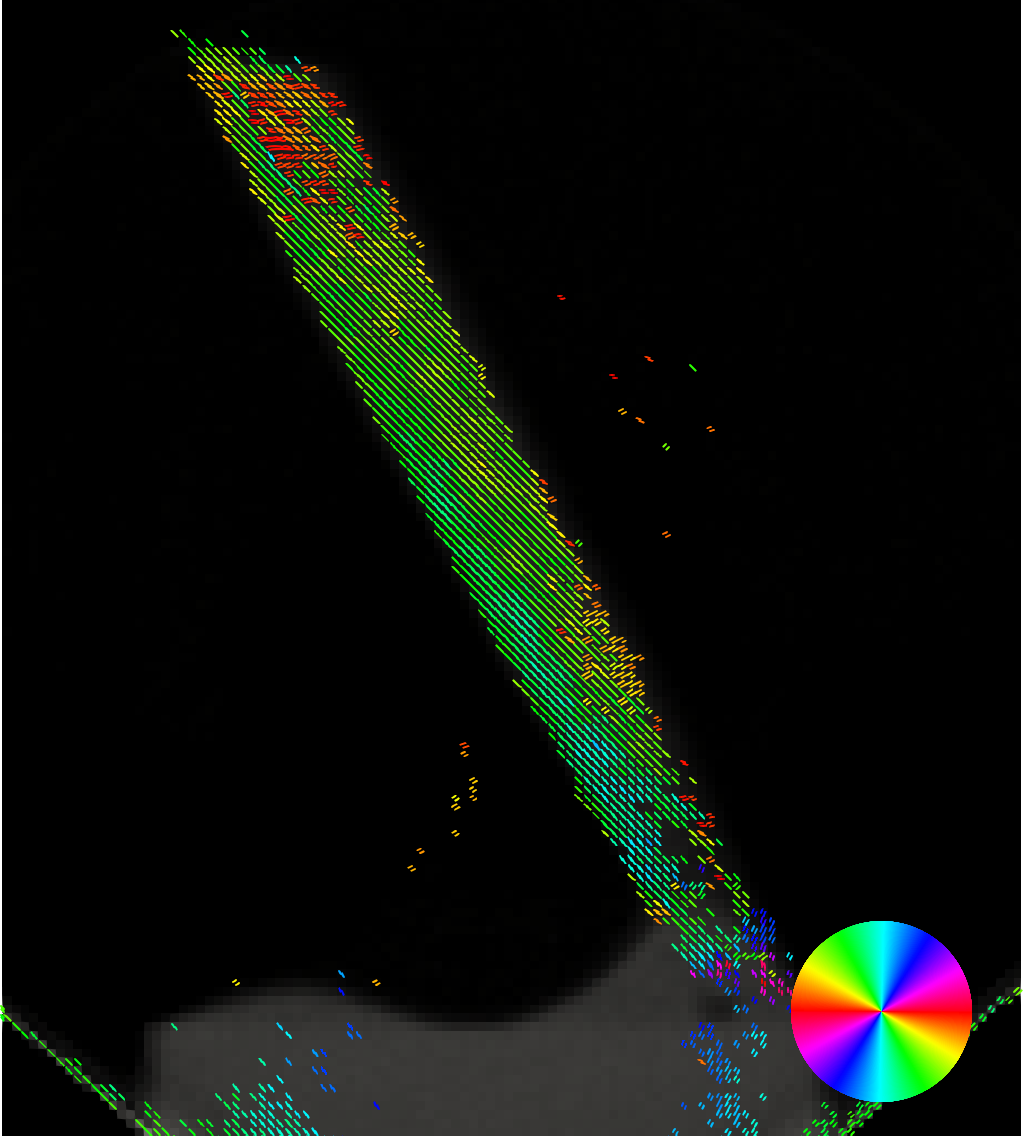}\hfil%
		\includegraphics[width=\linewidth]{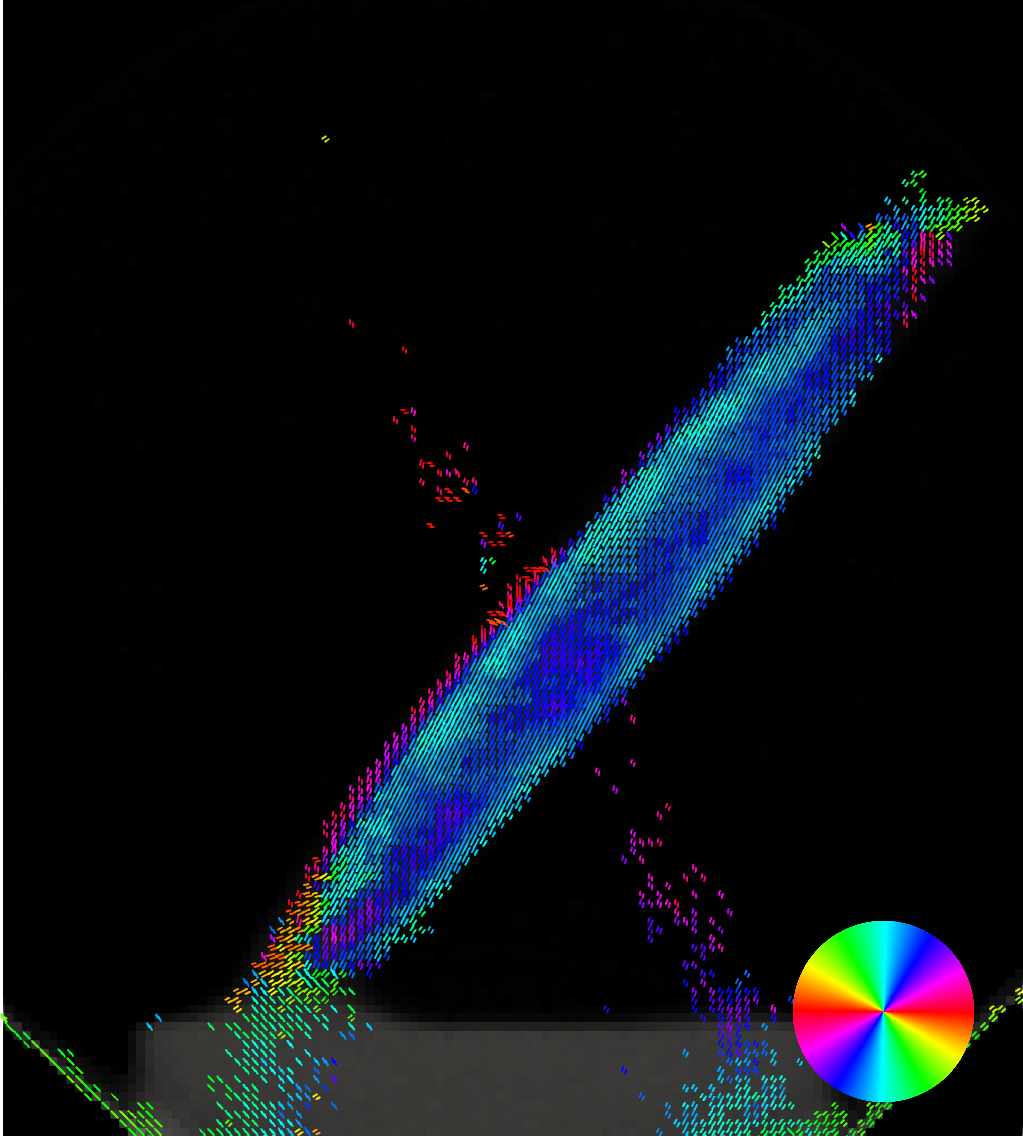}
		\caption{Visualization of model \textit{m2}}\label{fig:visu_sticks_m2}
	\end{subfigure}\hfil%
	\begin{subfigure}[t]{0.33\linewidth}
		\includegraphics[width=\linewidth]{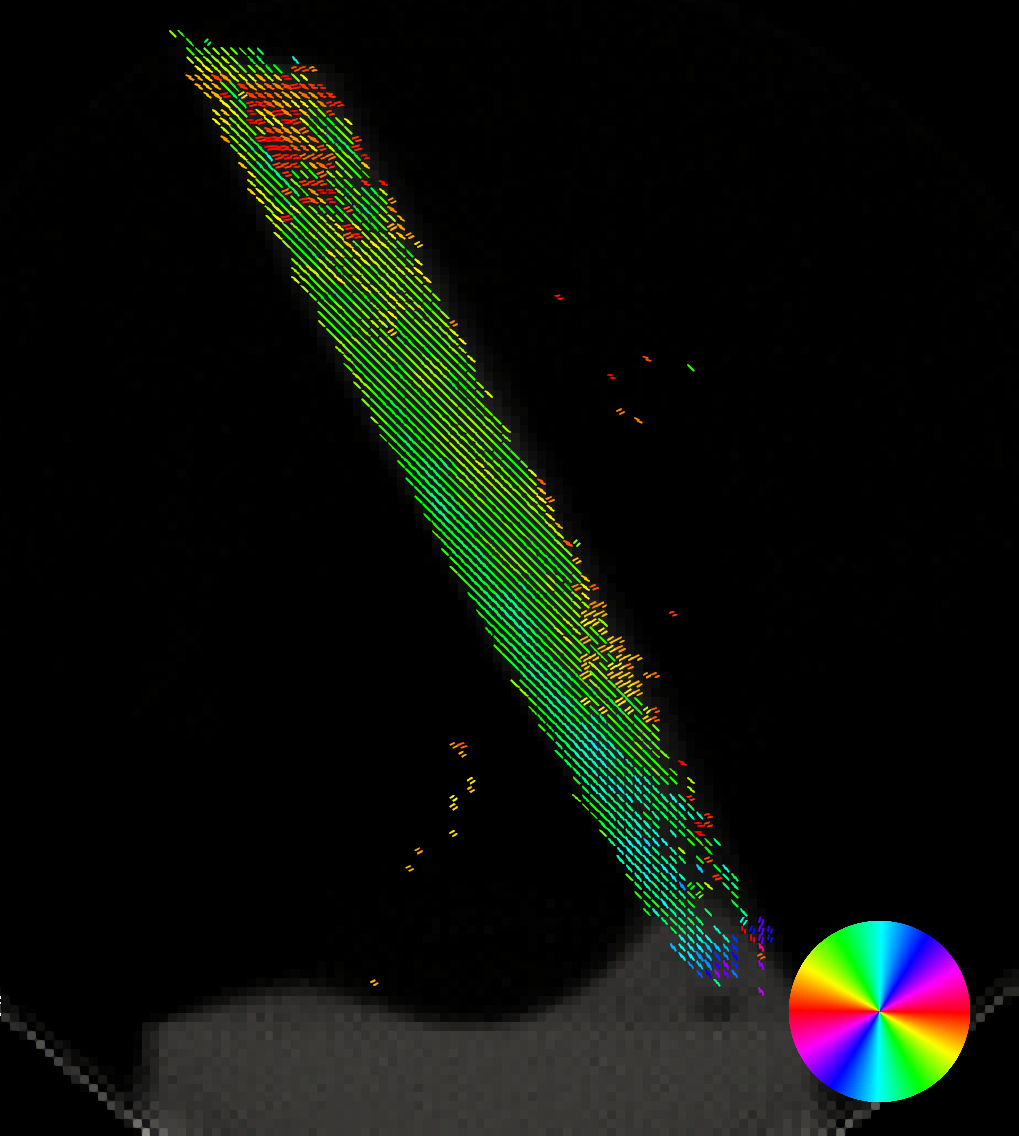}\hfil%
		\includegraphics[width=\linewidth]{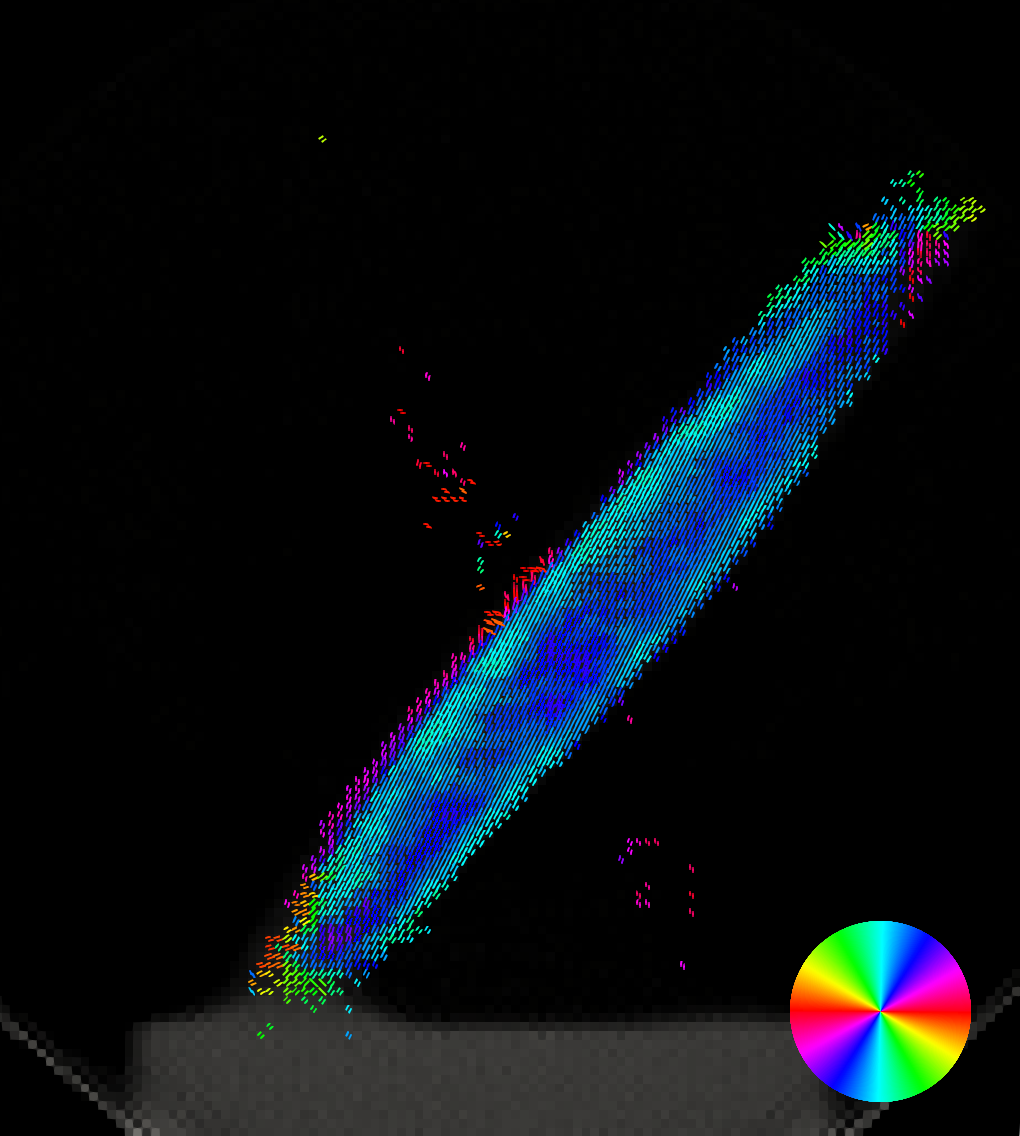}
		\caption{Visualization of model \textit{m3}}\label{fig:visu_sticks_m3}
	\end{subfigure}
	\caption{Fiber visualization of two slices from the crossed wooden sticks sample for each of the models overlaid on top of the attenuation X-ray CT reconstruction. The coloring shows the in-plane orientation according to the color wheel. Shown slices are xy-slices through each of the wooden sticks. From left to right, the visualization with the different models \textit{m1}, \textit{m2} and \textit{m3} is shown. Fibers are only shown where the scattering strength is above $0.00035$. Red arrows in visualization of model \textit{m1} indicate noticeable differences. We show visualization of the iterations \num{240} for both \textit{m1} and \textit{m3} and \num{1800} for \textit{m3}.}
	\label{fig:visu_sticks}
\end{figure}

\paragraph{Convergence Analysis} The convergence results from our experiments can be seen in Figure \ref{fig:convergence}. In the top row, the loss per iteration is plotted for each model. And in the bottom row, the loss over time is shown. The CG algorithm is the best choice for model \textit{m1}. It doesn't require any tweaking of parameters and performs basically as well as the L-BFGS algorithm, which requires the expensive Hessian for the line search. For the other models, however, L-BFGS reaches the lowest loss values in our experiments. This makes it particularly interesting for the advanced models compared to the non-linear CG algorithm. The performance of FGM is mostly dominated by the particular choice of the step length. If the parameter is well-chosen, it performs similar to the other algorithms and is particularly efficient, as no expensive Hessian must be evaluated for each step. The bounds for the Lipschitz constant we derived help choose good values quickly, but they still must be evaluated experimentally.

Table \ref{table:reco_time}, we give the required number of iterations and the reconstruction time for the visualizations down below. We want to draw particular attenuation to the performance of \textit{m3}. While still more expensive than model \textit{m1}, it vastly outperforms both the per iteration and overall required time compared to \textit{m2}. Unlike \textit{m1}, both statistical methods use a single iteration of Newton-Raphson's as line search and, as such, are expected to be computationally more expensive. From Figure \ref{fig:convergence}, we can see that an algorithm like FGM would substantially improve that runtime performance and result in a similar per iteration performance as the linear model. However, as the statistical models are only locally convex, FGM seems to struggle with monotonically decreasing the loss function.

\paragraph{Crossed Sticks Sample} In Figure \ref{fig:visu_sticks_harmonics}, we can see visualizations of the reconstruction of each model. In particular, we show the spherical harmonic coefficients $\eta_0^0$ and $\eta_2^1$. This is the direct output of the forward model. In the left row, we can see the reconstructions of model \textit{m1}; there, we can see strong noise in all the shown coefficients. $\eta_0^0$ clearly shows noise in both the inside of the stick, plus strong noise in the background. In the visualization of the coefficients of order \num{2}, rank \num{1}, the stick is barely noticeable due to the strong noise. The advanced models \textit{m2} and \textit{m3} improve this noise behavior noticeably. In the top row of Figure \ref{fig:visu_sticks_harmonics}, both statistical models improve the noise inside the stick and in the background. In the bottom row, the stick is now far better distinguishable from the background compared to model \textit{m1}.

The fiber extraction is capable of handling the noise level reconstructed from all models, as shown in Figure \ref{fig:visu_sticks}, where the fiber visualization of the different statistical noise models is shown. The visualization is restricted to the areas where the scattering strength is greater than $0.00035$. For \textit{m1} and \textit{m3}, we ran \num{280} iterations of CG and L-BFGS respectively, and \num{1800} iterations of L-BFGS for \textit{m2}. This again showcases the great convergence behavior of our proposed simplified statistical model. Choosing different threshold values resulted in either a noisy visualization or holes in the fiber visualization of model \textit{m1}. See the arrows labeled \num{1} and \num{2} in Figure \ref{fig:visu_sticks} The other models are easily thresholded to result in filled sticks. For both slices, one can also see that the models seen in the spherical-harmonic coefficients also carry over to the fiber extraction; see the arrow labeled \num{3}. The stick shown at the bottom of Figure \ref{fig:visu_sticks} seems particularly difficult to reconstruct. All methods require far more iterations to successfully show the desired fiber direction of \ang{45} for that stick compared to the one shown in the top row. The arrows \num{4}, \num{5} and \num{6} all point to areas that are fuzzier in model \textit{m1} than the statistical models.

\begin{figure}[!htbp]
	\centering
	\begin{subfigure}{0.33\linewidth}
		\includegraphics[width=\linewidth]{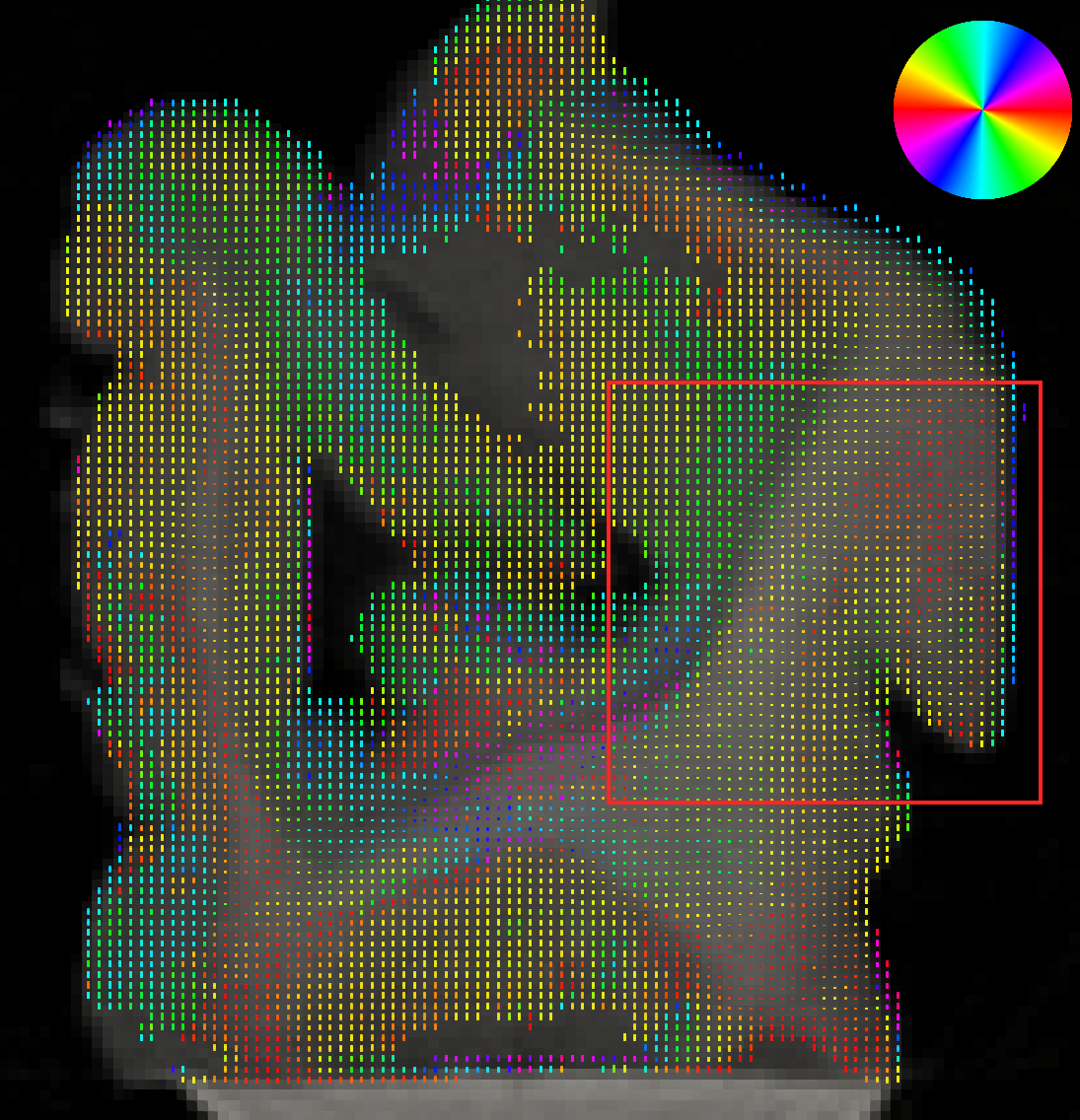}
		\includegraphics[width=\linewidth]{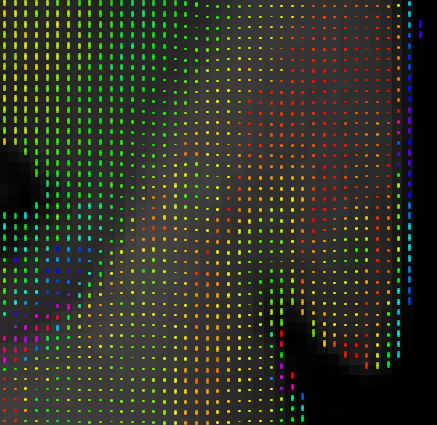}
		\caption{Visualization of model \textit{m1}}\label{fig:visu_brain_m0}
	\end{subfigure}\hfil%
	\begin{subfigure}{0.33\linewidth}
		\includegraphics[width=\linewidth]{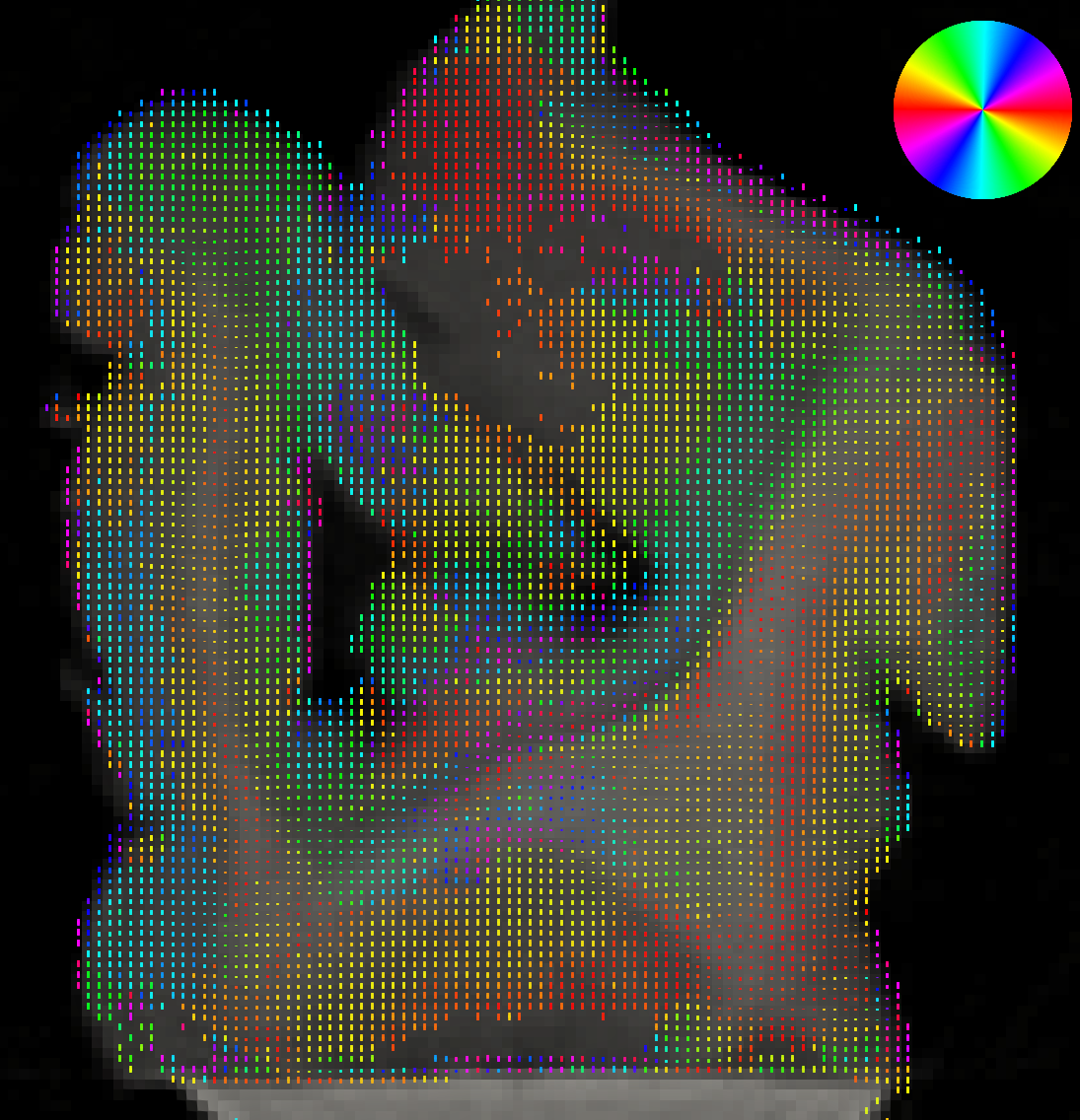}
		\includegraphics[width=\linewidth]{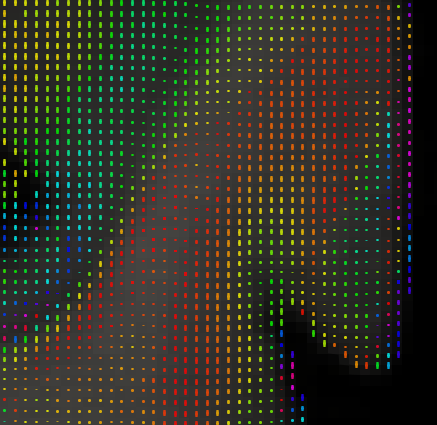}
		\caption{Visualization of model \textit{m2}}\label{fig:visu_brain_m2}
	\end{subfigure}\hfil%
	\begin{subfigure}{0.33\linewidth}
		\includegraphics[width=\linewidth]{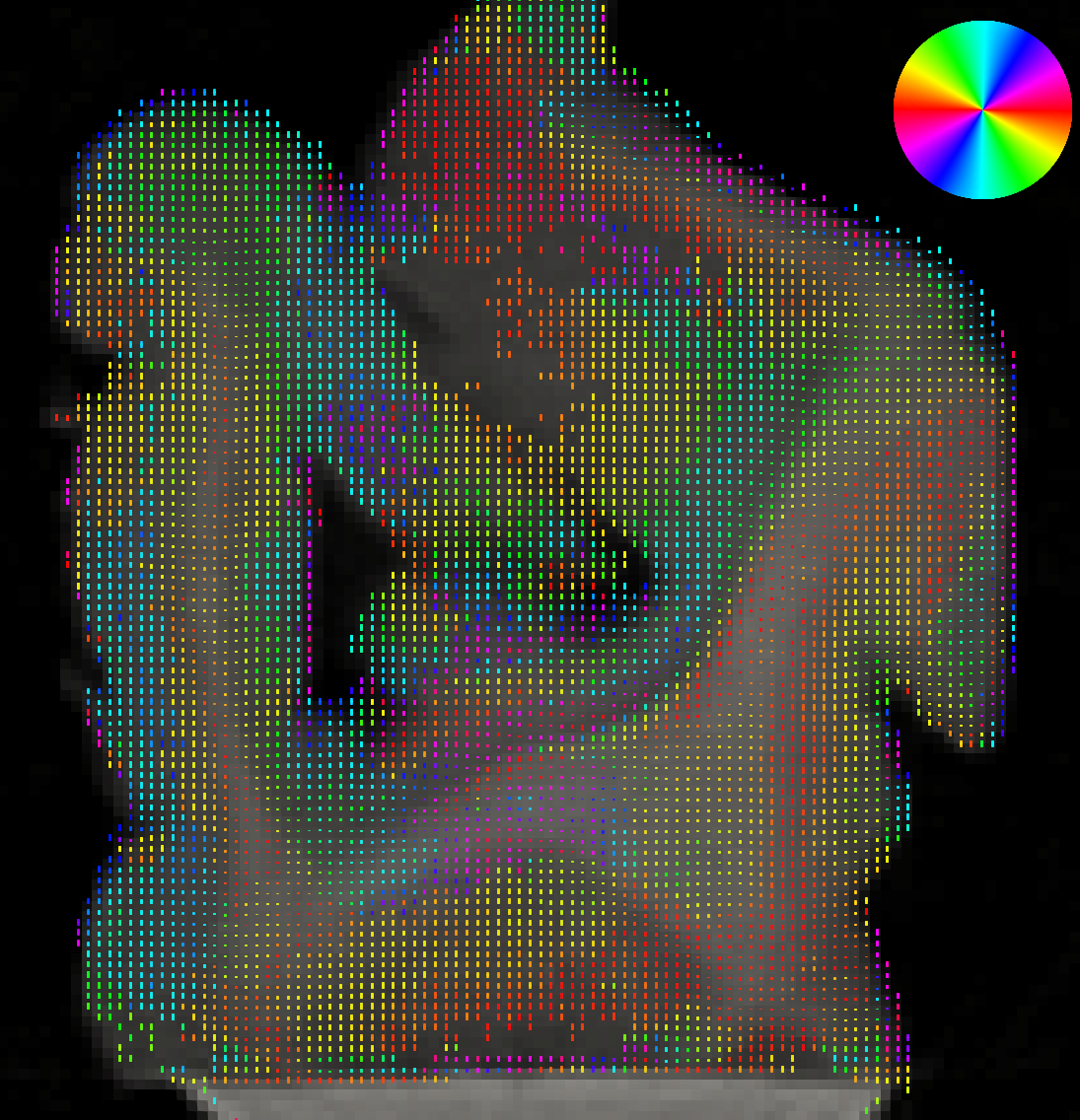}
		\includegraphics[width=\linewidth]{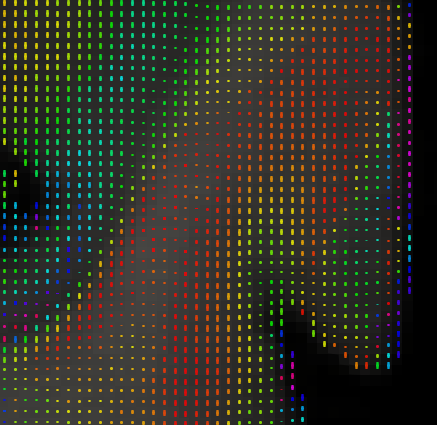}
		\caption{Visualization of model \textit{m3}}\label{fig:visu_brain_m3}
	\end{subfigure}
	\caption{Fiber visualization of the center slice from the human brain sample for each of the models overlaid on top of the attenuation X-ray CT reconstruction. The coloring shows the in-plane orientation according to the color wheel. From left to right, the visualization with the different models \textit{m1}, \textit{m2} and \textit{m3} is shown. The bottom row shows the crop indicated by the rectangle in the visualization of model \textit{m1}.}
	\label{fig:visu_brain}
\end{figure}

\paragraph{Brain Sample} In Figure \ref{fig:visu_brain}, the fiber extraction for the statistical reconstruction is given for the center slice of the human brain. The visualization is thresholded to values with scattering strength above $0.00015$. We see that the statistical models are clearly working even for complex and challenging biomedical datasets, such as this brain sample. While recognizing patterns in such a complex dataset without a verified ground truth is difficult, we want to draw attention to the reconstruction of structural information, as it can be seen in the cropped region shown at the bottom of Figure \ref{fig:visu_brain} While the linear model struggles to obtain any structural information in this part of the sample, the statistical models do show clear and distinguishable structure. Overall, the structural information is more difficult to spot in for model \textit{m1} compared to the other models. The required amount of iterations for all models was lower compared to the crossed sticks sample, as shown in Table \ref{table:reco_time}. Our proposed model again requires significantly less time to compute a comparable reconstruction to \textit{m2}.

\section{Conclusion}
\label{sec:conclusion}

As an emerging imaging technology, many open research problems remain for AXDT. Among others, an optimized reconstruction procedure can help decrease scanning time and, as such, reduce the required X-ray dose for a scan, in turn making it more accessible for further applications. Previous research used a noise model, which is not correct for the grating-based measurement system used for AXDT. In this paper, we provide a numerically stable implementation of previously proposed noise models, which shows improved reconstruction quality for two relevant datasets. The statistical reconstruction as previously proposed is however expensive to optimize, taking many iterations, and each iteration is expensive. As such, we propose a new propel formulation for AXDT, which builds on the actual noise assumption in GBI and provides similar reconstruction quality to the complex noise model but is computationally less expensive to compute and optimize. This improves the landscape of reconstruction methods for AXDT, by providing new methods that provide significant improvements in noise and image quality while still being feasible to reconstruct.

Further, we provide valuable insights into the mathematical properties of both the linearized and statistical models. In particular, we compute bounds in the Lipschitz constant, which provide useful for the explicit choice of reconstruction algorithm for statistical reconstruction of AXDT. We show that First-Order methods paired with our bounds on the Lipschitz constant are an extremely efficient tool for the reconstruction of AXDT. Further, we see that BFGS performs particularly well with the statistical reconstruction methods, providing excellent default convergence behavior without the need to many tune parameters. With this, we enable future research with a rich tool set of optimization techniques to further improve the reconstruction quality of AXDT.

\bibliographystyle{IEEEtran}
\bibliography{template}  

\end{document}